\documentclass[11pt,a4paper]{article}

\usepackage[margin=2.4cm]{geometry}
\usepackage[T1]{fontenc}
\usepackage[utf8]{inputenc}
\usepackage{mathptmx}
\usepackage{amsmath,amssymb,braket}
\usepackage{graphicx}
\usepackage{booktabs}
\usepackage{siunitx}
\usepackage{subcaption}
\usepackage[numbers,sort&compress]{natbib}
\usepackage{authblk}
\usepackage{xcolor}
\usepackage[colorlinks=true,linkcolor=blue!55!black,citecolor=blue!55!black,
            urlcolor=blue!55!black]{hyperref}
\usepackage{microtype}
\usepackage{caption}
\usepackage{titlesec}
\titleformat{\section}{\normalfont\large\bfseries}{\thesection}{0.7em}{}
\titleformat{\subsection}{\normalfont\normalsize\bfseries}{\thesubsection}{0.6em}{}

\newcommand{\Ntotal}{284{,}807}
\newcommand{\Nfraud}{492}
\newcommand{\BaseRatePct}{0.173}
\newcommand{\NqubitList}{2, 4, 8, 12}
\newcommand{\MaxQubits}{12}
\newcommand{\GridSize}{264}
\newcommand{\NDev}{4}
\newcommand{\NTest}{30}
\newcommand{\NFraudSample}{60}
\newcommand{\NLegitSample}{340}
\newcommand{\NSample}{400}
\newcommand{\EnrichPct}{15}
\newcommand{\FeatList}{\texttt{V14}, \texttt{V4}, \texttt{V11}, \texttt{V12}, \texttt{V10}, \texttt{V16}, \texttt{V3}, \texttt{V17}, \texttt{V9}, \texttt{V2}, \texttt{V7}, \texttt{V18}}
\newcommand{\QMeanTwo}{0.820}
\newcommand{\QSdTwo}{0.045}
\newcommand{\ProdMeanTwo}{0.807}
\newcommand{\ProdSdTwo}{0.056}
\newcommand{\ClsMeanTwo}{0.829}
\newcommand{\ClsSdTwo}{0.052}
\newcommand{\KMMeanTwo}{0.770}
\newcommand{\KMSdTwo}{0.061}
\newcommand{\SpecMeanTwo}{0.324}
\newcommand{\SpecSdTwo}{0.078}
\newcommand{\DiffQCTwo}{-0.0088}
\newcommand{\WinsQCTwo}{10}
\newcommand{\PQCTwo}{0.117}

\newcommand{\DiffQPTwo}{+0.0133}

\newcommand{\PQPTwo}{0.099}

\newcommand{\QMeanFour}{0.851}
\newcommand{\QSdFour}{0.040}
\newcommand{\ProdMeanFour}{0.860}
\newcommand{\ProdSdFour}{0.046}
\newcommand{\ClsMeanFour}{0.863}
\newcommand{\ClsSdFour}{0.043}
\newcommand{\KMMeanFour}{0.737}
\newcommand{\KMSdFour}{0.073}
\newcommand{\SpecMeanFour}{0.298}
\newcommand{\SpecSdFour}{0.151}
\newcommand{\DiffQCFour}{-0.0119}
\newcommand{\WinsQCFour}{10}
\newcommand{\PQCFour}{0.025}

\newcommand{\QMeanEight}{0.867}
\newcommand{\QSdEight}{0.041}
\newcommand{\ProdMeanEight}{0.855}
\newcommand{\ProdSdEight}{0.041}
\newcommand{\ClsMeanEight}{0.858}
\newcommand{\ClsSdEight}{0.046}
\newcommand{\KMMeanEight}{0.686}
\newcommand{\KMSdEight}{0.083}
\newcommand{\SpecMeanEight}{0.244}
\newcommand{\SpecSdEight}{0.103}
\newcommand{\DiffQCEight}{+0.0086}
\newcommand{\WinsQCEight}{19}
\newcommand{\PQCEight}{0.013}
\newcommand{\CIQCEight}{[+0.0020, +0.0153]}
\newcommand{\DiffQPEight}{+0.0118}

\newcommand{\PQPEight}{0.004}
\newcommand{\BestFamEight}{angle-ent}
\newcommand{\BestTopoEight}{alternating}

\newcommand{\BestEntEight}{14}

\newcommand{\NClsConfigsEight}{105}
\newcommand{\BwMinEight}{0.707}
\newcommand{\BwMaxEight}{0.870}
\newcommand{\BwRangeEight}{0.163}
\newcommand{\QMeanTwelve}{0.855}
\newcommand{\QSdTwelve}{0.045}
\newcommand{\ProdMeanTwelve}{0.846}
\newcommand{\ProdSdTwelve}{0.045}
\newcommand{\ClsMeanTwelve}{0.852}
\newcommand{\ClsSdTwelve}{0.044}
\newcommand{\KMMeanTwelve}{0.637}
\newcommand{\KMSdTwelve}{0.103}
\newcommand{\SpecMeanTwelve}{0.236}
\newcommand{\SpecSdTwelve}{0.083}
\newcommand{\DiffQCTwelve}{+0.0026}
\newcommand{\WinsQCTwelve}{12}
\newcommand{\PQCTwelve}{0.422}

\newcommand{\DiffQPTwelve}{+0.0090}

\newcommand{\PQPTwelve}{0.013}

\newcommand{\MaxAbsDiff}{0.0119}
\newcommand{\BwVsDiff}{14}

\newcommand{\BudgetMax}{264}

\newcommand{\BudSmall}{20}
\newcommand{\BudSmallMargin}{-0.0017}
\newcommand{\BudSmallPos}{34}

\newcommand{\BrQAriHi}{0.842}
\newcommand{\BrCAriHi}{0.836}

\newcommand{\BrQAriLo}{0.003}
\newcommand{\BrCAriLo}{0.000}

\newcommand{\TrPct}{0.10}
\newcommand{\TrN}{2002}
\newcommand{\TrQAuc}{0.949}
\newcommand{\TrCAuc}{0.955}
\newcommand{\TrDAuc}{0.949}

\newcommand{\ShotPoints}{150}
\newcommand{\ShotCircuits}{11{,}175}

\newcommand{\ShotAriEightSOneZeroTwoFour}{0.893}

\newcommand{\ShotExactEight}{0.893}

\newcommand{\ShotAriTwelveSOneZeroTwoFour}{0.846}

\newcommand{\ShotExactTwelve}{0.846}
\newcommand{\ConcLoN}{2}
\newcommand{\ConcHiN}{14}
\newcommand{\ConcOffLo}{0.80}
\newcommand{\ConcOffHi}{0.51}
\newcommand{\ConcGapLo}{0.59}
\newcommand{\ConcGapHi}{0.49}
\newcommand{\ConcSnrLo}{1.85}
\newcommand{\ConcSnrHi}{1.63}
\newcommand{\ConcAriLo}{0.85}
\newcommand{\ConcAriHi}{0.81}

\newcommand{\ari}{\mathrm{ARI}}

\title{\bfseries Quantum Kernel $k$-Means for Credit-Card Fraud Detection:\\
A Controlled Benchmark on Real Transaction Data}

\author[1]{Muhammad Faryad\thanks{Correspondence: \texttt{muhammad.faryad@lums.edu.pk}}}
\affil[1]{\small Department of Physics, Lahore University of Management Sciences, Lahore, Pakistan}

\date{\today}

\begin{document}
\maketitle

\begin{abstract}
\noindent
We benchmarked quantum kernel $k$-means against classical clustering for
credit-card fraud detection on real transaction data, at up to \MaxQubits{}
qubits, under a protocol with separated selection and reporting data and
matched search budgets. We find no robust quantum advantage: the sign of the
difference depends on register size, all effect sizes are below $0.013$ ARI,
and the single significant advantage we observe is fully explained by the
number of configurations searched. We further show that ordinary
hyperparameter choices move performance by considerably more than the quantum
kernel does, that additional qubits degrade rather than improve performance
through kernel concentration, and that the clustering framing itself fails at
realistic class imbalance though kernel-based anomaly scoring does not.
We regard the methodological contribution as the more durable one. The
search-budget ablation in particular is inexpensive and, in our case, decisive:
it converted a statistically significant advantage into a procedural artefact.
We would encourage its routine use.
\end{abstract}


\section{Introduction}

The identification of a near-term quantum advantage for machine learning is an
open problem of considerable practical interest. Quantum kernel methods
\citep{havlicek2019,schuld2019} are a leading candidate: they require only
state preparation and a measurement of state overlap, place the optimisation
loop on classical hardware, and are supported by the observation that
essentially all supervised quantum models with fixed encodings are kernel
methods \citep{schuld2021kernel}. Their appeal is that a quantum computer
supplies a similarity function that may be hard to evaluate classically, while
the surrounding algorithm remains familiar.

Credit-card fraud detection recurs as a motivating application. It is
commercially valuable, labels are expensive and arrive late, and the data are
already released in a form---principal components of confidential
fields---that is convenient for angle encoding. It is therefore a natural
setting in which to ask whether a quantum kernel does anything useful.

The difficulty is that most reported comparisons are not designed to be able
to fail. Three problems recur. First, the classical baseline is often plain
$k$-means or an untuned kernel, whereas the quantum model is the best of many
circuits the authors examined; the comparison then measures search effort
rather than physics. Second, model selection and reporting are performed on
the same data, so the reported score includes the selection luck. Third,
single-split results are quoted without a variance estimate, which for
clustering on resampled subsets is large enough to swamp the effects being
claimed. Recent work has drawn attention to precisely these failure modes
\citep{bowles2024benchmarking,gilfuster2024generalization}.

This paper reports a benchmark constructed to avoid them. Our contributions
are:

\begin{enumerate}\itemsep2pt
\item A \textbf{budget-matched, held-out protocol} for quantum-versus-classical
comparison: model selection on development resamples, reporting on \NTest{}
disjoint held-out resamples, and an explicitly matched number of classical
configurations (Sec.~\ref{sec:protocol}).
\item A \textbf{register-size study} of quantum kernel $k$-means on real fraud
data at \NqubitList{} qubits, over \GridSize{} feature-map configurations
spanning seven families and four entanglement topologies
(Sec.~\ref{sec:scaling}).
\item A \textbf{search-budget ablation} that quantifies how much of an apparent
quantum advantage is purchased by trying more circuits
(Sec.~\ref{sec:budget}). To our knowledge this control has not previously
been reported for quantum kernel methods, and it fully accounts for the one
statistically significant advantage we observe.
\item A \textbf{mechanistic account} of why additional qubits do not help,
based on direct measurement of kernel concentration
(Sec.~\ref{sec:concentration}).
\item A \textbf{practical reframing}: clustering fails for all methods at
realistic fraud base rates, whereas kernel-based anomaly scoring does not---but
the quantum kernel offers no advantage there either
(Sec.~\ref{sec:baserate}).
\end{enumerate}

Our conclusion is negative but, we argue, useful: on this task, under matched
conditions, quantum kernels are \emph{competitive with} and not superior to
tuned classical kernels, and the margin in either direction is far smaller
than the effect of ordinary hyperparameter choices.

\section{Background}

\subsection{Quantum feature maps and fidelity kernels}

A feature map is a parameterised circuit $U(\mathbf{x})$ that prepares
$\ket{\psi_\mathbf{x}} = U(\mathbf{x})\ket{0}^{\otimes n}$ from a classical
vector $\mathbf{x}\in\mathbb{R}^n$. The induced \emph{fidelity kernel} is
\begin{equation}
K(\mathbf{x},\mathbf{y}) \;=\;
\big|\braket{\psi_\mathbf{y}|\psi_\mathbf{x}}\big|^{2}
\;=\; \mathrm{Tr}\!\left[\rho_\mathbf{x}\rho_\mathbf{y}\right],
\label{eq:kernel}
\end{equation}
with $\rho_\mathbf{x} = \ket{\psi_\mathbf{x}}\!\bra{\psi_\mathbf{x}}$. This is
a positive semi-definite kernel: it is the Hilbert--Schmidt inner product
between density matrices, so any kernel method may consume it directly.

\subsection{Kernel \texorpdfstring{$k$}{k}-means}

Lloyd's algorithm \citep{lloyd1982} alternates assignment and centroid update.
Centroids in the quantum feature space cannot be materialised as data points,
so we use the kernel formulation \citep{dhillon2004kernel}, in which the
centroid of cluster $C$ remains implicit and only kernel evaluations appear:
\begin{equation}
\big\|\rho_\mathbf{x} - \mathbf{m}_C\big\|^{2}
= K(\mathbf{x},\mathbf{x})
- \frac{2}{|C|}\sum_{i\in C} K(\mathbf{x},\mathbf{x}_i)
+ \frac{1}{|C|^{2}}\sum_{i,j\in C} K(\mathbf{x}_i,\mathbf{x}_j).
\label{eq:kkm}
\end{equation}
Since $K(\mathbf{x},\mathbf{x}) = 1$ for pure states, the first term is
constant and does not affect the assignment. The quantum device is therefore
used exactly once, to populate the $N\times N$ Gram matrix; everything
afterwards is classical.

\subsection{Measuring the overlap}

Equation~\eqref{eq:kernel} is not directly observable. Two circuits estimate
it. The SWAP test \citep{buhrman2001fingerprinting} uses an ancilla and
controlled-SWAP gates on two registers, requiring $2n+1$ qubits and returning
$P(\text{ancilla}=0) = (1+K)/2$.

We instead use the \emph{inversion} (compute--uncompute) test, which requires
only $n$ qubits. Applying $U(\mathbf{x})$ and then $U(\mathbf{y})^\dagger$ and
measuring all qubits in the computational basis gives
\begin{equation}
P(0^{\otimes n})
= \big|\bra{0^{\otimes n}} U(\mathbf{y})^\dagger U(\mathbf{x})
  \ket{0^{\otimes n}}\big|^{2}
= \big|\braket{\psi_\mathbf{y}|\psi_\mathbf{x}}\big|^{2}
= K(\mathbf{x},\mathbf{y}),
\label{eq:inversion}
\end{equation}
so the kernel entry is read directly off the all-zeros count with no
post-processing. Besides halving the qubit count, this avoids the factor of
two in $K = 2P-1$ that amplifies sampling error in the SWAP test.

\section{Data and protocol}
\label{sec:protocol}

\subsection{Dataset}

We use the ULB credit-card fraud dataset \citep{dalpozzolo2015}, comprising
\Ntotal{} transactions recorded over two days in September 2013, of which
\Nfraud{} (\BaseRatePct\%) are fraudulent. Features \texttt{V1}--\texttt{V28}
are principal components of the original confidential fields; only
\texttt{Time} and \texttt{Amount} are untransformed. We obtain the data from
OpenML (ID 1597).

\subsection{Feature selection and encoding}

We rank the 28 components by standardised mean difference between classes
(Fig.~\ref{fig:features}) and take the top $n$ for an $n$-qubit experiment,
giving \FeatList{} for $n=\MaxQubits$. Features are standardised and then
min--max scaled to $[0,\pi]$; a bandwidth $c$ multiplies the scaled values
before encoding, following \citet{shaydulin2022bandwidth}.

\subsection{Class balance}

At a fraud rate of \BaseRatePct\% no two-cluster method can isolate the
minority class; this is a property of the task, not of any algorithm. As is
standard for clustering studies on this dataset, we therefore work on
\emph{enriched} resamples of \NFraudSample{} frauds and \NLegitSample{}
legitimate transactions (\EnrichPct\% fraud, $N=\NSample$). This is not a
realistic detection setting, and we quantify the cost of the enrichment
explicitly in Sec.~\ref{sec:baserate} rather than leaving it implicit.

\subsection{Evaluation metric}

Accuracy is unusable under imbalance: on our resamples a degenerate assignment
placing every point in one cluster scores $0.85$. We therefore report the
adjusted Rand index \citep{hubert1985comparing}, which counts pair-wise
agreements between two partitions and subtracts the expectation under a
random model, so that $\ari=1$ is exact recovery, $\ari=0$ is chance, and
negative values are worse than chance.

\subsection{Selection and reporting protocol}

The protocol is the methodological core of this work.

\begin{itemize}\itemsep2pt
\item \textbf{Development resamples} (seeds $0$--$\the\numexpr\NDev-1\relax$)
      are used \emph{only} to choose hyperparameters, for both the quantum and
      the classical model.
\item \textbf{Held-out resamples} (\NTest{} independent draws, seeds
      $100$--$129$) are used \emph{only} for reporting, and are never
      consulted during selection.
\item \textbf{Matched search budgets.} The quantum side searches \GridSize{}
      feature-map configurations. The classical side searches
      \NClsConfigsEight{} kernel configurations spanning radial basis
      function, Laplacian, polynomial and sigmoid kernels with dense bandwidth
      grids.
\item \textbf{Paired statistics.} Because every method sees the same
      resamples, we report paired $t$-tests and $95\%$ confidence intervals on
      the per-resample difference, not comparisons of independent means.
\end{itemize}

\subsection{Feature-map search space}

We search seven feature-map families---plain angle encoding; angle encoding
with a CNOT layer; the ZZ map of \citet{havlicek2019}; an IQP-style map; a
data re-uploading map \citep{perezsalinas2020reupload}; a hardware-efficient
map; and a mixed-basis Pauli map---crossed with four entanglement topologies
(none, linear, circular, and brickwork), one and two repetitions, and six
bandwidths $c \in \{0.25, 0.5, 0.75, 1.0, 1.5, 2.0\}$, giving \GridSize{}
configurations. We exclude all-to-all connectivity, which requires
$\mathcal{O}(n^2)$ two-qubit gates and is not representative of near-term
hardware \citep{preskill2018}.

\subsection{Simulation}

All circuits are simulated with Qiskit Aer \citep{qiskit2024}. For the
large-scale searches we evaluate Eq.~\eqref{eq:kernel} exactly from
statevectors; Sec.~\ref{sec:shots} repeats the full pipeline with kernels
estimated from finite measurement samples and confirms that the conclusions
are unchanged. Classical baselines use scikit-learn \citep{scikitlearn}.

\section{Results}

\begin{figure}[t]
\centering
\includegraphics[width=0.72\textwidth]{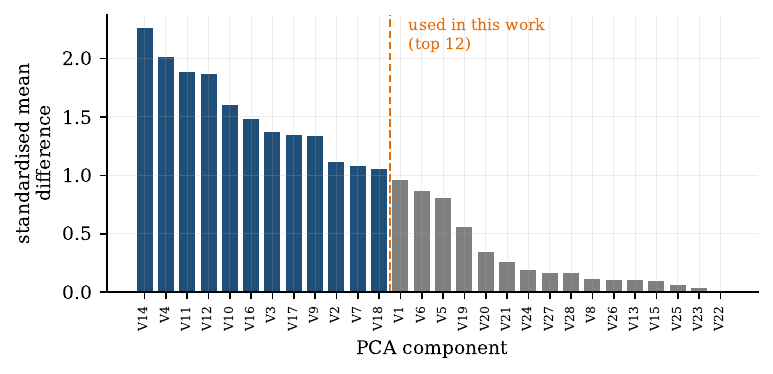}
\caption{Class separation of the 28 principal components, measured as the
standardised mean difference between fraudulent and legitimate transactions.
The $n$-qubit experiments use the top $n$ components; separation falls off
sharply beyond the first eight.}
\label{fig:features}
\end{figure}

\subsection{Register size does not buy an advantage}
\label{sec:scaling}

\begin{figure}[t]
\centering
\includegraphics[width=\textwidth]{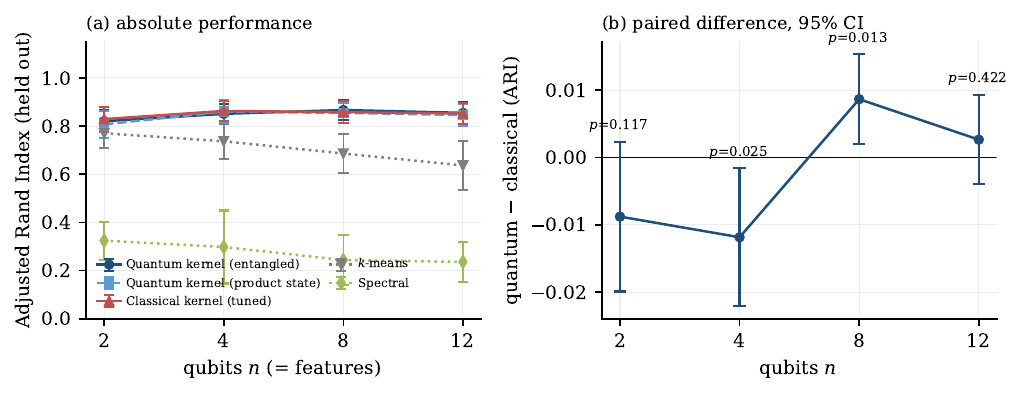}
\caption{(a) Held-out adjusted Rand index against register size for all
methods; error bars are standard deviations over \NTest{} resamples.
(b) Paired quantum-minus-classical difference with $95\%$ confidence
intervals. The sign of the difference changes with register size and its
magnitude never exceeds $0.012$.}
\label{fig:scaling}
\end{figure}

Figure~\ref{fig:scaling} and Table~\ref{tab:main} report held-out performance
at \NqubitList{} qubits. Three observations stand out.

First, all kernel methods---quantum and classical---clearly outperform plain
$k$-means, which reaches only $\KMMeanEight \pm \KMSdEight$ at eight qubits.
Spectral clustering performs poorly throughout. The interesting comparison is
therefore between kernels, not between quantum and $k$-means.

Second, the quantum kernel is \emph{not} uniformly better or worse. At two and
four qubits it is behind the tuned classical kernel, significantly so at four
($\Delta\ari = \DiffQCFour$, $p = \PQCFour$). At eight qubits it is ahead
($\Delta\ari = \DiffQCEight$, $p = \PQCEight$, 95\% CI $\CIQCEight$). At
twelve the difference is not distinguishable from zero
($\Delta\ari = \DiffQCTwelve$, $p = \PQCTwelve$).

Third, and most importantly, \emph{every} difference is small. The largest
magnitude in either direction is $0.012$ ARI, on resamples whose
between-draw standard deviation is around $0.04$. Absolute performance peaks at
eight qubits ($\QMeanEight$) and falls at twelve ($\QMeanTwelve$), so the
common expectation that a larger Hilbert space should help is not borne out.

\begin{table}[t]
\centering
\small
\caption{Held-out adjusted Rand index (mean $\pm$ s.d.\ over \NTest{}
independent resamples). Quantum and classical kernels are each selected on
development resamples only, with matched search budgets. $\Delta$ is the mean
paired difference between the quantum kernel and the tuned classical kernel;
$p$ is from a paired $t$-test.}
\label{tab:main}
\begin{tabular}{lcccc}
\toprule
& \multicolumn{4}{c}{register size $n$}\\
\cmidrule(lr){2-5}
Method & 2 & 4 & 8 & 12 \\
\midrule
$k$-means                   & $\KMMeanTwo \pm \KMSdTwo$ & $\KMMeanFour \pm \KMSdFour$ & $\KMMeanEight \pm \KMSdEight$ & $\KMMeanTwelve \pm \KMSdTwelve$\\
Spectral clustering         & $\SpecMeanTwo \pm \SpecSdTwo$ & $\SpecMeanFour \pm \SpecSdFour$ & $\SpecMeanEight \pm \SpecSdEight$ & $\SpecMeanTwelve \pm \SpecSdTwelve$\\
Quantum kernel, product     & $\ProdMeanTwo \pm \ProdSdTwo$ & $\ProdMeanFour \pm \ProdSdFour$ & $\ProdMeanEight \pm \ProdSdEight$ & $\ProdMeanTwelve \pm \ProdSdTwelve$\\
Classical kernel, tuned     & $\ClsMeanTwo \pm \ClsSdTwo$ & $\ClsMeanFour \pm \ClsSdFour$ & $\ClsMeanEight \pm \ClsSdEight$ & $\ClsMeanTwelve \pm \ClsSdTwelve$\\
Quantum kernel, entangled   & $\QMeanTwo \pm \QSdTwo$ & $\QMeanFour \pm \QSdFour$ & $\mathbf{\QMeanEight \pm \QSdEight}$ & $\QMeanTwelve \pm \QSdTwelve$\\
\midrule
$\Delta$ (quantum $-$ classical) & $\DiffQCTwo$ & $\DiffQCFour$ & $\DiffQCEight$ & $\DiffQCTwelve$\\
$p$                              & $\PQCTwo$ & $\PQCFour$ & $\PQCEight$ & $\PQCTwelve$\\
wins / \NTest{}                  & $\WinsQCTwo$ & $\WinsQCFour$ & $\WinsQCEight$ & $\WinsQCTwelve$\\
\bottomrule
\end{tabular}
\end{table}

The configuration selected on development data is remarkably stable: at every
register size the winner is angle encoding followed by a CNOT layer, with two
repetitions and bandwidth $c=1$ (\BestFamEight{}, \BestTopoEight{} topology,
\BestEntEight{} entangling gates at $n=8$). The classical winner is a
Laplacian kernel at $n \geq 4$.

\subsection{Bandwidth matters more than the choice of family}

\begin{figure}[t]
\centering
\includegraphics[width=\textwidth]{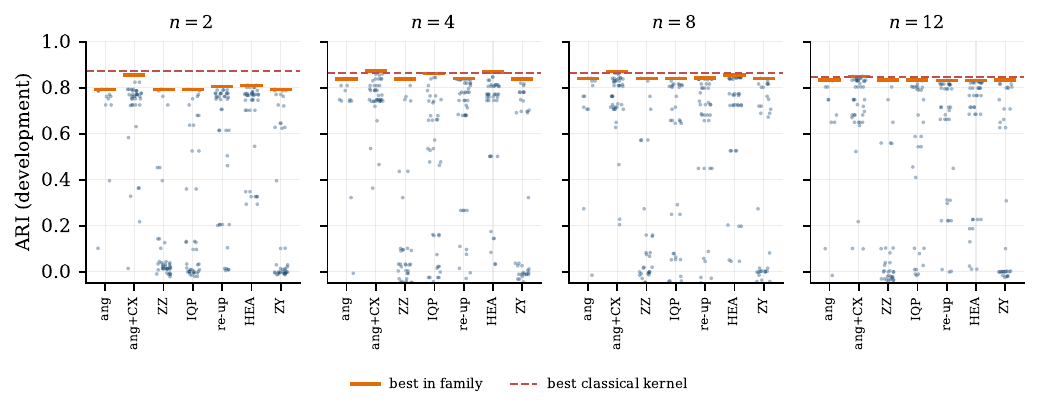}
\caption{Development-set adjusted Rand index for all \GridSize{} feature-map
configurations at each register size, grouped by family (one point per
configuration, jittered horizontally). Orange bars mark the best configuration
in each family; the dashed red line is the best classical kernel. Variation
\emph{within} a family exceeds variation between families.}
\label{fig:search}
\end{figure}

Figure~\ref{fig:search} shows the development score of all \GridSize{}
configurations, grouped by family. Within-family spread is far larger than
between-family spread: the best member of almost every family reaches a
similar score, while the worst members are close to zero. Restricting to the
best configuration per bandwidth
(Fig.~\ref{fig:bw}a), performance varies from \BwMinEight{} to \BwMaxEight{}
at eight qubits---a range of \BwRangeEight{}, roughly $\BwVsDiff\times$ the
largest quantum-versus-classical difference we measure ($\MaxAbsDiff$). This corroborates
\citet{shaydulin2022bandwidth} and \citet{canatar2023bandwidth} in an applied
setting and has a direct practical consequence: a quantum kernel benchmark
that does not report its bandwidth has not reported its model.

\subsection{Entanglement helps, slightly, and only when the register is large enough}

\begin{figure}[t]
\centering
\includegraphics[width=\textwidth]{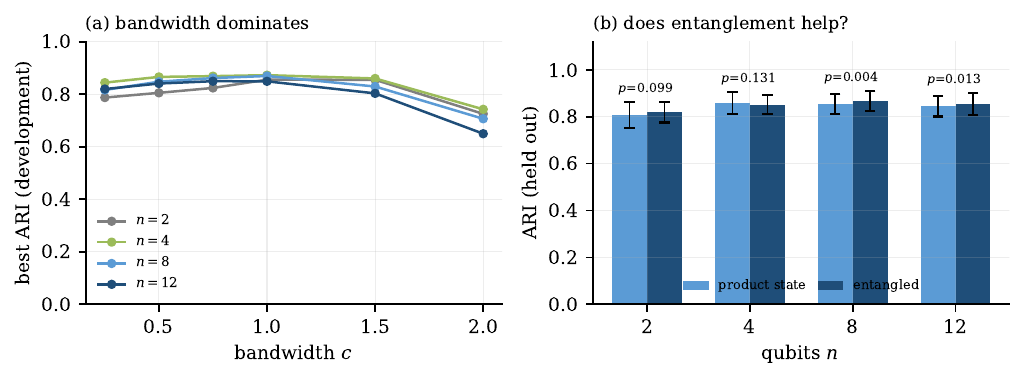}
\caption{(a) Best development score as a function of encoding bandwidth $c$.
A single hyperparameter spans a far wider range than any
quantum-versus-classical difference reported here. (b) Held-out comparison of
the best entangling map against the best product-state map, each tuned
independently.}
\label{fig:bw}
\end{figure}

Comparing the best entangling map with the best product-state map on held-out
data (Fig.~\ref{fig:bw}b) shows a consistent but modest benefit that grows with
register size: $\DiffQPTwo$ at two qubits ($p = \PQPTwo$) against
$\DiffQPEight$ at eight ($p = \PQPEight$) and $\DiffQPTwelve$ at twelve
($p = \PQPTwelve$).

A methodological caution accompanies this result. If entangling gates are
simply removed without re-optimising the bandwidth---the natural way to run
the ablation---the apparent benefit is inflated several-fold, because the
bandwidth was tuned for the entangled circuit. Each arm of such an ablation
must be tuned separately; we report only separately-tuned comparisons.

\subsection{The eight-qubit advantage is a search-budget artefact}
\label{sec:budget}

The one statistically significant advantage in Table~\ref{tab:main} was
obtained by selecting the best of \GridSize{} configurations. We therefore ask
how the held-out margin depends on how many configurations the quantum side is
permitted to examine. We score every configuration on all held-out resamples,
then repeatedly draw random subsets of size $B$, select within each subset
using development data only, and record the resulting held-out margin.

\begin{figure}[t]
\centering
\includegraphics[width=\textwidth]{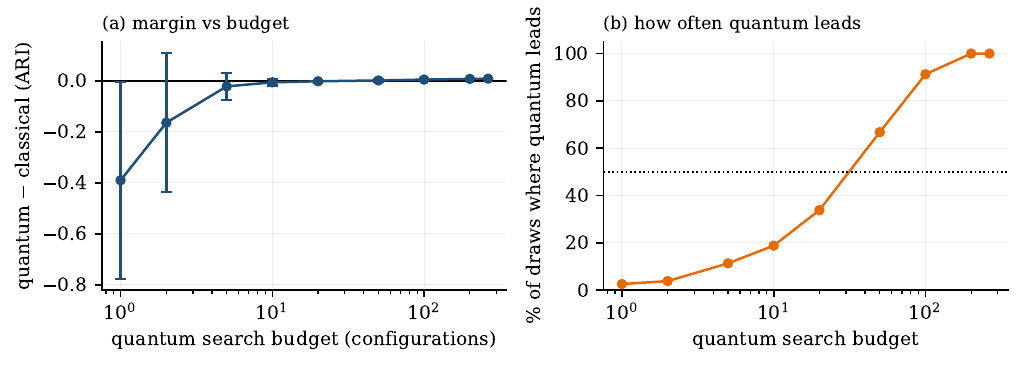}
\caption{Search-budget ablation at eight qubits. (a) Mean held-out
quantum-minus-classical margin as a function of the number of quantum
configurations from which the model is selected (400 random draws per budget;
error bars are standard deviations across draws). (b) Fraction of draws in
which the quantum kernel leads. The apparent advantage grows monotonically
with search effort and is negative at small budgets.}
\label{fig:budget}
\end{figure}

Figure~\ref{fig:budget} shows the result. The margin is a monotone function of
the search budget. With $B \leq \BudSmall$ the expected margin is negative
(\BudSmallMargin{}) and the quantum kernel leads in only \BudSmallPos\% of
draws; it crosses zero at roughly $B \approx 50$; and only at the full budget
does it reach the \DiffQCEight{} reported above, at which point it leads in
$100\%$ of draws by construction.

The advantage is therefore purchased, not discovered. Since the classical side
was allowed \NClsConfigsEight{} configurations, the honest reading of
Table~\ref{tab:main} is that at comparable budgets the two are indistinguishable.
We suggest that reporting this curve, or at minimum the number of quantum
configurations evaluated, should become routine.

\subsection{Why more qubits do not help: kernel concentration}
\label{sec:concentration}

\begin{figure}[t]
\centering
\includegraphics[width=\textwidth]{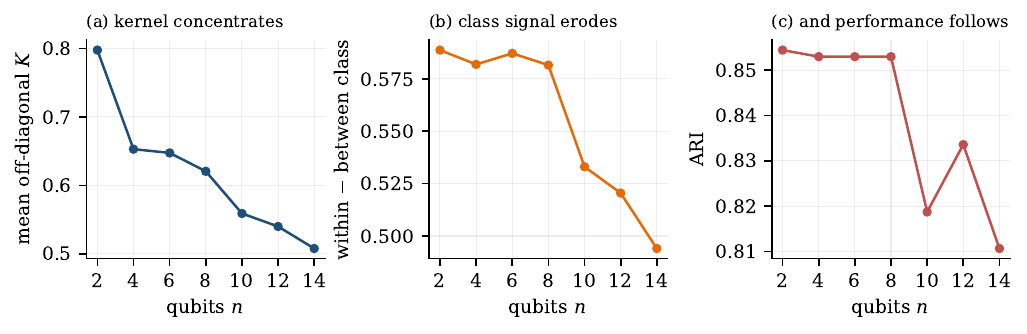}
\caption{Kernel concentration with register size, for the selected feature map
held fixed. (a) The mean off-diagonal kernel entry falls as the register
grows. (b) The gap between mean within-class and mean between-class kernel
values---the signal the clustering actually uses---erodes. (c) Adjusted Rand
index follows. Both intrinsic concentration and the declining class
information of the added components (Fig.~\ref{fig:features}) contribute.}
\label{fig:conc}
\end{figure}

Fidelity kernels are expected to concentrate as the Hilbert-space dimension
grows, with off-diagonal entries collapsing toward a constant and the Gram
matrix approaching the identity \citep{thanasilp2024concentration}. We measure
this directly (Fig.~\ref{fig:conc}). Holding the selected map fixed and
increasing the register from \ConcLoN{} to \ConcHiN{} qubits, the mean
off-diagonal kernel value falls from $\ConcOffLo$ to $\ConcOffHi$, and---more
consequentially---the gap between the mean within-class and mean
between-class kernel value falls from $\ConcGapLo$ to $\ConcGapHi$, so the
signal-to-noise ratio of the Gram matrix degrades from $\ConcSnrLo$ to
$\ConcSnrHi$. Adjusted Rand index falls correspondingly from $\ConcAriLo$ to
$\ConcAriHi$.

Two effects are entangled here and we separate them explicitly. Part of the
degradation is concentration in the sense of
\citet{thanasilp2024concentration}. Part is simply that the components added
beyond the first eight carry much less class information
(Fig.~\ref{fig:features}), so extra qubits encode mostly noise. Both point the
same way: for this dataset there is no benefit in going beyond roughly eight
qubits, and the limiting factor is the data, not the simulator.

\subsection{Finite sampling on the Aer simulator}
\label{sec:shots}

The results above use exact statevector overlaps. To confirm that they survive
realistic measurement, we repeat the full pipeline with every kernel entry
estimated from finite measurement samples on Aer, using
Eq.~\eqref{eq:inversion} with \ShotPoints{} points
(\ShotCircuits{} circuits per Gram matrix).

\begin{figure}[t]
\centering
\includegraphics[width=\textwidth]{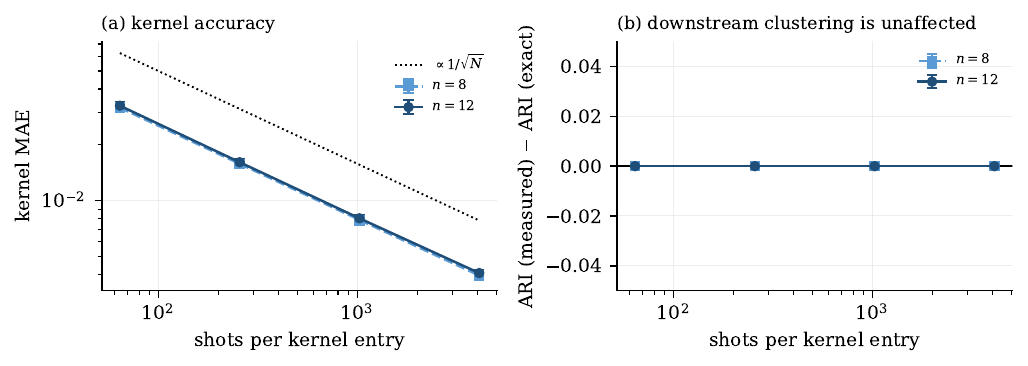}
\caption{Full pipeline with kernels measured on the Aer simulator via the
inversion test. (a) Mean absolute error of the Gram matrix against shot count,
following the expected $N^{-1/2}$ scaling. (b) Difference between the
clustering obtained from the measured Gram matrix and from the exact one; it
is indistinguishable from zero at every shot count tested. Kernel $k$-means
depends on the Gram matrix only through cluster-averaged quantities, so
independent per-entry noise largely averages out.}
\label{fig:shots}
\end{figure}

Figure~\ref{fig:shots} shows the mean absolute kernel error falling as
$N^{-1/2}$ as expected for a binomial estimator, and---more usefully---that
downstream clustering is remarkably tolerant of this noise. At eight qubits
the measured pipeline reproduces the exact adjusted Rand index
($\ShotAriEightSOneZeroTwoFour$ at $1024$ shots against $\ShotExactEight$
exact); at twelve qubits the same holds
($\ShotAriTwelveSOneZeroTwoFour$ against $\ShotExactTwelve$). Kernel
$k$-means depends on the Gram matrix only through cluster-averaged quantities,
so independent per-entry noise largely averages out.

This is encouraging for hardware feasibility but does not change the
comparison: the sampling cost is $N(N-1)/2$ circuit evaluations per Gram
matrix, which at $N=10^4$ is $5\times10^{7}$ circuits, each requiring hundreds
of shots.

\subsection{Class imbalance, and a better framing of the task}
\label{sec:baserate}

\begin{figure}[t]
\centering
\includegraphics[width=\textwidth]{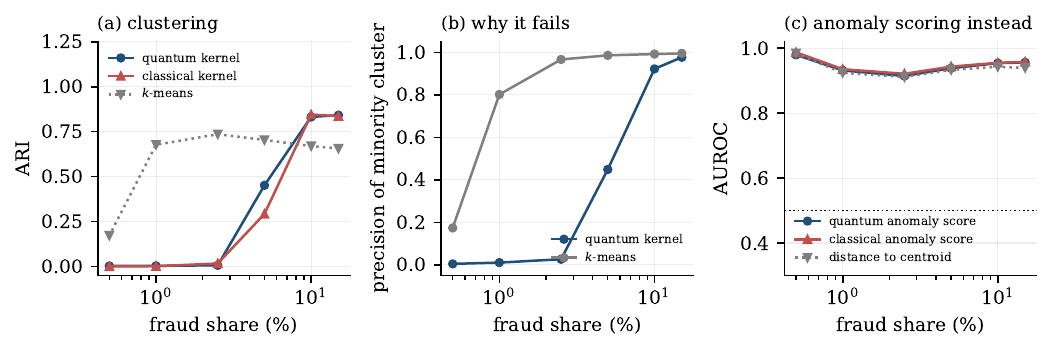}
\caption{Behaviour as class imbalance approaches the true base rate.
(a) Clustering quality; both kernel methods collapse below about $5\%$ fraud
while $k$-means degrades gracefully. (b) Precision of the minority cluster,
showing the mechanism. (c) The same kernels used for anomaly \emph{scoring}
rather than clustering retain high AUROC throughout---but the quantum kernel
matches, and does not beat, a classical kernel or simple distance to the
centroid.}
\label{fig:baserate}
\end{figure}

All results so far use \EnrichPct\% fraud. Figure~\ref{fig:baserate}a shows
what happens as the fraud share is reduced toward its true value. Both kernel
methods collapse below about $5\%$: adjusted Rand index falls from
$\BrQAriHi$ to $\BrQAriLo$ for the quantum kernel and from $\BrCAriHi$ to
$\BrCAriLo$ for the classical one. Plain $k$-means degrades far more
gracefully. The mechanism is visible in Fig.~\ref{fig:baserate}b: the
precision of the minority cluster collapses while recall does not. A kernel
with a single global bandwidth cannot form a tight cluster around a handful of
extreme outliers, and instead absorbs ordinary points.

This is a limitation of kernel clustering, not of quantum computing, and it
suggests the task has been framed wrongly. Fraud detection is an anomaly
detection problem. We therefore score each point by its mean kernel similarity
to all other points and rank by that score, flagging the least similar. This
requires the same Gram matrix and no clustering at all.
Figure~\ref{fig:baserate}c shows that this framing survives the imbalance:
at a $\TrPct$\% fraud rate ($N=\TrN$) the quantum kernel reaches AUROC
$\TrQAuc$. However, the classical kernel reaches $\TrCAuc$ and plain distance
to the centroid reaches $\TrDAuc$. The reframing helps; the quantum kernel
does not.

\section{Discussion}

Taken together the results support a simple claim: on this dataset, under
matched search budgets and honest held-out evaluation, quantum kernel
$k$-means is competitive with tuned classical kernel clustering and is not
better than it.

We emphasise what this does \emph{not} say. It is not evidence that quantum
kernels can never help; the search space of feature maps is vast and we
explored a structured but finite corner of it. It is not a statement about
other datasets; the ULB features are principal components, which may be
particularly well matched to classical shift-invariant kernels. And it is not
a statement about asymptotic complexity, since we simulate exactly and ignore
any potential speedup in evaluating the kernel itself.

What it does say is that the effect sizes typically reported in this
literature are of the same order as, or smaller than, the effects of routine
methodological choices. We measured three such choices, each of which moved
performance by more than the quantum--classical difference: the encoding
bandwidth (up to \BwRangeEight{} ARI), the number of configurations searched
(Sec.~\ref{sec:budget}), and whether both arms of an ablation were tuned. A
benchmark that does not control all three cannot distinguish a physical effect
from a procedural one.

The register-size result deserves particular emphasis because it runs against
a common intuition. Moving from 8 to 12 qubits enlarges the Hilbert space
sixteen-fold and \emph{reduces} performance. Part of this is intrinsic
concentration of fidelity kernels
\citep{thanasilp2024concentration,kubler2021inductive}; part is that the extra
features carry little signal. Either way, the number of qubits is not the
resource that is scarce here.

\subsection{Recommendations for applied quantum machine learning benchmarks}

We suggest four practices, all cheap to adopt:
\begin{enumerate}\itemsep2pt
\item Report the number of quantum configurations evaluated, and give the
classical baseline a comparable budget.
\item Separate selection data from reporting data, and report paired
statistics with confidence intervals across multiple resamples.
\item Report the encoding bandwidth and its sensitivity.
\item When ablating a component, re-tune the remaining hyperparameters in each
arm.
\end{enumerate}

\section{Limitations}

Our search space, while broad, excludes trainable feature maps and
all-to-all connectivity. We simulate noiseless circuits; hardware noise would
degrade the quantum side only, so our conclusions are conservative in that
direction. The enrichment to \EnrichPct\% fraud is unrealistic, which we
address directly in Sec.~\ref{sec:baserate} but cannot eliminate for the
clustering task. Our $t$-tests treat resamples as independent although they are
drawn from a common pool; the non-parametric Wilcoxon test gives the same
conclusions. Finally, we use one dataset: the results should not be
extrapolated to problems with different geometry, particularly ones where the
data are believed to have quantum-native structure \citep{huang2021power}.

\section{Conclusion}

Quantum kernel methods are among the most frequently proposed near-term
applications of quantum computing to machine learning, and fraud detection is
among the most frequently cited use cases. Reported advantages, however, are
usually established against weakly tuned classical baselines and without
controlling for the number of quantum model configurations examined. We
benchmark quantum kernel $k$-means against classical clustering on the
real ULB credit-card fraud dataset (\Ntotal{} transactions, \Nfraud{}
frauds), at register sizes of \NqubitList{} qubits, under a protocol designed
to make the comparison falsifiable: model selection is performed on
development resamples only, all results are reported on \NTest{} held-out
resamples never used for selection, and the classical baseline receives a
search budget matched to the quantum one. Overlaps are evaluated with the
inversion (compute--uncompute) test on the Qiskit Aer simulator. Across
\GridSize{} quantum feature-map configurations per register size we find no
robust advantage. The sign of the quantum--classical difference changes with
register size: the quantum kernel is significantly \emph{worse} at four qubits
($\Delta\ari = \DiffQCFour$, $p = \PQCFour$) and significantly better at eight
($\Delta\ari = \DiffQCEight$, $p = \PQCEight$), but the difference vanishes at
twelve ($\Delta\ari = \DiffQCTwelve$, $p = \PQCTwelve$). A search-budget
ablation shows the eight-qubit advantage to be an artefact of selection: when
the quantum model is chosen from \BudSmall{} configurations rather than
\BudgetMax{}, the mean margin is \BudSmallMargin{} and quantum leads in only
\BudSmallPos{}\% of draws. All effect sizes are below $0.013$ adjusted Rand
index in either direction, while a single classical preprocessing
hyperparameter---the encoding bandwidth---moves performance by up to
\BwRangeEight{}. We further show that no method of either kind clusters fraud
successfully at realistic class imbalance, that recasting the task as
kernel-based anomaly scoring restores useful performance
(AUROC $\TrQAuc$ at a $\TrPct$\% fraud rate), and that this too is matched by
classical kernels and by simple distance to the centroid. We argue that
budget-matched, held-out benchmarking should be a minimum standard for applied
quantum machine learning claims, and we release all code and data splits.

\section*{Code and data availability}

The dataset is publicly available from OpenML (ID 1597) and Kaggle. All code
required to reproduce the experiments, figures and tables, including the exact
resample seeds used for selection and reporting, is available at
\url{https://github.com/muf148/quantum-kmeans-fraud}.

\section*{Acknowledgements}
We thank the participants of the \emph{Unsupervised Quantum Machine Learning}
workshop, whose questions motivated the search-budget ablation reported in
Sec.~\ref{sec:budget}.

\bibliographystyle{unsrtnat}
\bibliography{refs}

\end{document}